\documentclass[10pt,letterpaper]{article}
\usepackage[margin=1in]{geometry}
\usepackage{parskip}
\usepackage{amsmath,amsfonts,amssymb}
\usepackage{graphicx}
\usepackage{array}
\usepackage{hyperref}
\usepackage[font=small]{caption}
\usepackage{quotes}
\usepackage{booktabs}
\usepackage[table]{xcolor}
\usepackage{lmodern}
\usepackage{authblk}
\usepackage{titlesec}
\usepackage[numbers, sort&compress]{natbib}
\usepackage{float}
\usepackage{abstract}

\titlespacing*{\section}{0pt}{2.0ex plus 1ex minus .2ex}{1.0ex}
\titlespacing*{\subsection}{0pt}{1.5ex plus 1ex minus .2ex}{0.8ex}

\titleformat{\section}
  {\centering\normalfont\large\bfseries}{\thesection}{1em}{}

\titleformat{\subsection}
  {\centering\normalfont\normalsize\bfseries}{\thesubsection}{1em}{}

\titleformat{\subsubsection}
  {\centering\normalfont\normalsize\itshape}{\thesubsubsection}{1em}{}

\titleformat{\title}
  {\normalfont\Large\bfseries}{\thetitle}{1em}{}

\definecolor{goodblue}{RGB}{0, 91, 187}
\definecolor{color1}{RGB}{45, 93, 131}
\definecolor{color2}{RGB}{79, 155, 217}
\definecolor{color3}{RGB}{7, 113, 135}
\definecolor{color4}{RGB}{61, 250, 255}
\definecolor{color5}{RGB}{139, 136, 142}
\hypersetup{
  colorlinks=true,
  allcolors=color2,
  urlcolor=color2,
  citecolor=color2,
  pdfborder={0 0 0},
  breaklinks=true,
}

\newcommand{\xhdr}[1]{\vspace{0.1in}\noindent\textbf{#1}}
\title{\normalsize\textbf{A Blue Start: A large-scale pairwise and higher-order social network dataset}}

\author[1]{Alyssa Smith\thanks{smith.alyss@northeastern.edu}}
\author[2]{Ilya Amburg\thanks{ilya.amburg@pnnl.gov}}
\author[1,3]{Sagar Kumar\thanks{kumar.sag@northeastern.edu}}
\author[1,4]{Brooke Foucault Welles\thanks{b.welles@northeastern.edu}}
\author[5,6,7]{Nicholas W. Landry\thanks{nicholas.landry@virginia.edu}}

\affil[1]{Network Science Institute, Northeastern University, Boston, Massachusetts, USA}
\affil[2]{Pacific Northwest National Laboratory, Seattle, Washington, USA}
\affil[3]{Center for Health Informatics, Boston Children's Hospital, Boston, Massachusetts, USA}
\affil[4]{Department of Communication Studies, Northeastern University, Boston, Massachusetts, USA}
\affil[5]{Department of Biology, University of Virginia, Charlottesville, Virginia, USA}
\affil[6]{School of Data Science, University of Virginia, Charlottesville, Virginia, USA}
\affil[7]{Vermont Complex Systems Institute, University of Vermont, Burlington, Vermont, USA}
\date{{\small \today}}

\begin{document}
\raggedbottom
\maketitle

\begin{abstract}
Large-scale networks have been instrumental in shaping how we think about social systems, and have undergirded many foundational results in mathematical epidemiology, computational social science, and biology.
However, many of the social systems through which diseases spread, information disseminates, and individuals interact are inherently mediated through groups, known as higher-order interactions.
A gap exists between higher-order models of group formation and spreading processes and the data necessary to validate these mechanisms.
Similarly, few datasets bridge the gap between pairwise and higher-order network data.
The Bluesky social media platform is an ideal laboratory for observing social ties at scale through its open API.
Not only does Bluesky contain pairwise following relationships, but it also contains higher-order social ties known as "starter packs" which are user-curated lists designed to promote social network growth.
We introduce "A Blue Start", a large-scale network dataset comprising 39.7M user accounts, 2.4B pairwise following relationships, and 365.8K groups representing starter packs.
This dataset will be an essential resource for the study of higher-order networks.
\end{abstract}

\setlength{\parskip}{3pt}
\section*{Background \& Summary}

Network science has revolutionized the study of human behavior through its ability to represent complex relationships between individuals in a population.
Representing social structures as networks offers insights on the temporal nature of interactions \cite{cattuto_dynamics_2010}, the heterogeneity of interaction patterns \cite{ebel_scalefree_2002,barabasi_evolution_2002,gonzalez_understanding_2008}, preferential relationships between nodes with similar characteristics \cite{newman_assortative_2002,newman_mixing_2003} and highly connected nodes \cite{vaquero_rich_2013}, and community structure \cite{girvan_community_2002}.
One particular medium that spawned an entire sub-field of network science was Twitter, the social media platform.
This online "laboratory" was particularly useful because of its open automated programming interface (API); the free "Twitter Streaming API" provided researchers with up to 1\% of all tweets, while the paid "Twitter Firehose" allowed researchers to extract 100\% of all public posts~\cite{morstatter_sample_2013}.
As Twitter grew in size, the Firehose API was discontinued and replaced with the Decahose API which allowed researchers to download 10\% of all posts in real time~\cite{campan_data_2018,li_how_2016}.
The relative availability of Twitter data, along with the platform's popularity and use across demographic and sociographic groups~\cite{mislove_understanding_2011} spurred scientific innovations across a number of disciplines, including epidemic modeling~\cite{ginsberg_detecting_2009}, political communication~\cite{kreiss_seizing_2016}, and statistical physics~\cite{gleeson_effects_2016}, to name just a few. Some sub-disciplines that depend on access to "everyday" communication, such as social movement studies, were revolutionized by access to the Twitter API, which provided data to support hundreds of papers on the structure and discursive content of social movements around the world (e.g., see Refs.~\cite{gerbaudo_tweets_2012,jackson_hashtagactivism_2020,tufekci_twitter_2017}). 

The change in ownership of Twitter, subsequently known as X, spawned interest in many alternative platforms such as Mastodon, Threads, and Bluesky~\cite{li_crossplatform_2024,quelle_why_2025,seckin_rise_2025}.
We focus our attention of Bluesky because of its inherent open access to platform data.
Bluesky originated in 2019 as a research initiative at Twitter examining the potential for open-sourcing and decentralizing social media~\cite{palmer_twitter_2019}.
It was incorporated as an independent company in 2021 ~\cite{mccue_how_2024}.
Bluesky's central innovation is a standard known as the Authenticated Transfer (AT) Protocol~\cite{kleppmann_bluesky_2024}.
The AT Protocol emphasizes federated identities, allowing users to own and curate a single online identity that is usable on all AT Protocol platforms and services~\cite{blueskypbc_federation_2025}.
User data is stored in repositories on personal data servers (PDSs) that log user actions; these repositories are intended to be portable so that users can easily migrate between hosts~\cite{blueskypbc_repository_2025}.
The Bluesky \emph{application} queries a user's PDS to deliver content to users; however, anyone can build an application to view AT Protocol data.
Applications with access to the same data can vary in their content moderation policies, post visibility, or algorithmic curation strategies, for example~\cite{blueskyteam_moderation_2023}.

Although large-scale social network datasets derived from online social media platforms such as Twitter~\cite{morstatter_sample_2013}, Bluesky~\cite{jeong_descriptor_2024,failla_im_2024,balduf_looking_2024,balduf_bootstrapping_2025}, Facebook~\cite{bond_61millionperson_2012}, and many others~\cite{zhang_characterizing_2014, takac_data_2012, backstrom_group_2006, mahoney_community_2009, mislove_measurement_2007, rossi_network_2015, kunegis_konect_2013, leskovec_snap_2016, fu_modelling_2021} have formed the experimental basis of many studies on human behavior, these studies almost exclusively analyze these networks from a pairwise perspective, since following relationships are inherently pairwise.
In contrast, many social systems contain interactions which may not be pairwise but rather may involve an arbitrary number of individuals.
Higher-order networks are composed of the collection of these interactions, known as \textit{higher-order interactions}, and can more accurately represent the structure of many empirical systems~\cite{landry_filtering_2024,landry_simpliciality_2024,chodrow_configuration_2020}.
For example, in the scientific ecosystems, grants and scientific publications often involve large teams of collaborators, and the size~\cite{guimera_team_2005,wuchty_increasing_2007,wu_large_2019}, interdisciplinarity~\cite{shi_surprising_2023}, and composition~\cite{chowdhary_team_2024} of these teams can be linked to scientific outcomes such as citations and collaboration trajectories.
However, co-authorship networks have historically been represented as pairwise networks~\cite{newman_coauthorship_2004}, where a publication with \textit{k} authors is converted to a $k$-clique, making it challenging to disambiguate, for example, a paper authored by A, B, and C, or three papers authored by A and B, B ad C, and A and C.
Similarly, in computational social science, higher-order networks can help model the formation and fragmentation of groups \cite{iacopini_temporal_2024}, face-to-face interactions~\cite{gallo_higherorder_2024}, and network structure at different scales of interaction~\cite{landry_filtering_2024}.
There are relatively few \textit{truly} higher-order social network datasets, however.
Of those that do exist, they can largely be categorized as email datasets~\cite{benson_simplicial_2018} and proximity datasets~\cite{stehle_highresolution_2011,cattuto_dynamics_2010,mastrandrea_contact_2015,vanhems_estimating_2013} for which maximal simultaneous cliques of pairwise proximity measurements have been converted to higher-order interactions~\cite{benson_simplicial_2018}.
In fact, several statistical approaches have been developed to infer higher-order interactions from pairwise networks \cite{benson_simplicial_2018,young_hypergraph_2021,lizotte_hypergraph_2023}.
However, there is a lack of both intrinsically higher-order social network datasets and datasets comprising both higher-order interactions and pairwise interactions.

This Bluesky dataset addresses both of these gaps in the corpus of higher-order network datasets.
In the summer of 2024, Bluesky developers announced a new feature in the Bluesky app, known as \textit{starter packs} (SPs).
Starter packs are user-curated lists of accounts and feeds (which serve post content to users based on custom algorithms) with a user-supplied title and description.
Users share these lists so that other users can quickly grow the list of accounts they follow based on specified interest.
One has the option to either browse the list and cherry-pick accounts to follow or hit the "follow all" button to follow all accounts in the list.
Starter packs are inherently higher-order.
At the time of this writing, users can create starter packs comprising 8 to 150 users and feeds.
One caveat, however, is that a similar higher-order structure on Bluesky known as "lists", which are not size restricted and so contain anywhere between zero and several thousand items.
With the advent of starter packs, users developed tools for converting lists into starter packs, bypassing the standard size limits.
So while the list of starter packs is almost entirely higher-order, there are exceptions to this (as can be later seen in Fig.~\ref{fig:sp_stats}).
One can also extract the list of accounts that each user follows, which can be represented as a directed, pairwise network.
These two types of social relationships bridge the gap between pairwise and higher-order network data.
Furthermore, this dataset offers researchers ground to investigate social phenomena exclusive to groups and develop measures and algorithms tailored to such structures.
In particular, the absence of edges in the following network when compared to the network induced by starter packs facilitates exploration of patterns that are exclusive to higher-order networks.
For example, one can analyze user overlap patterns between starter packs, a structural feature inaccessible to pairwise network analysis.
Higher-order network analysis enables a deeper understanding of the complex dynamics among users participating in group interactions.

Beyond applications for the study of higher-order dynamics and interactions, our data enables novel research in communication and human-computer interaction.
Future work in this area can leverage our dataset to address questions of structural differences in starter pack inclusion, the effect of starter pack inclusion on following relationships, and more.
This dataset provides an ideal testbed for developing machine learning algorithms which use pairwise features to learn higher-order network structure and vice-versa.
Engineering systems to solve these tasks can unlock new insights into the relationship between peer-to-peer social networks and group-level interactions online.
This dataset can also supplement the suite of higher-order social datasets used to validate group formation mechanisms and simulate the higher-order dynamics of information, norms, and social capital.
Historical online social network datasets have been used to answer diverse research questions across a wide range of disciplines and our Bluesky dataset adds to this rich corpus, providing a unique opportunity to gain early insights into a key, distinguishing feature of a burgeoning new social media platform.

\section*{Methods}

\begin{figure}
    \centering
    \includegraphics[width=0.75\linewidth]{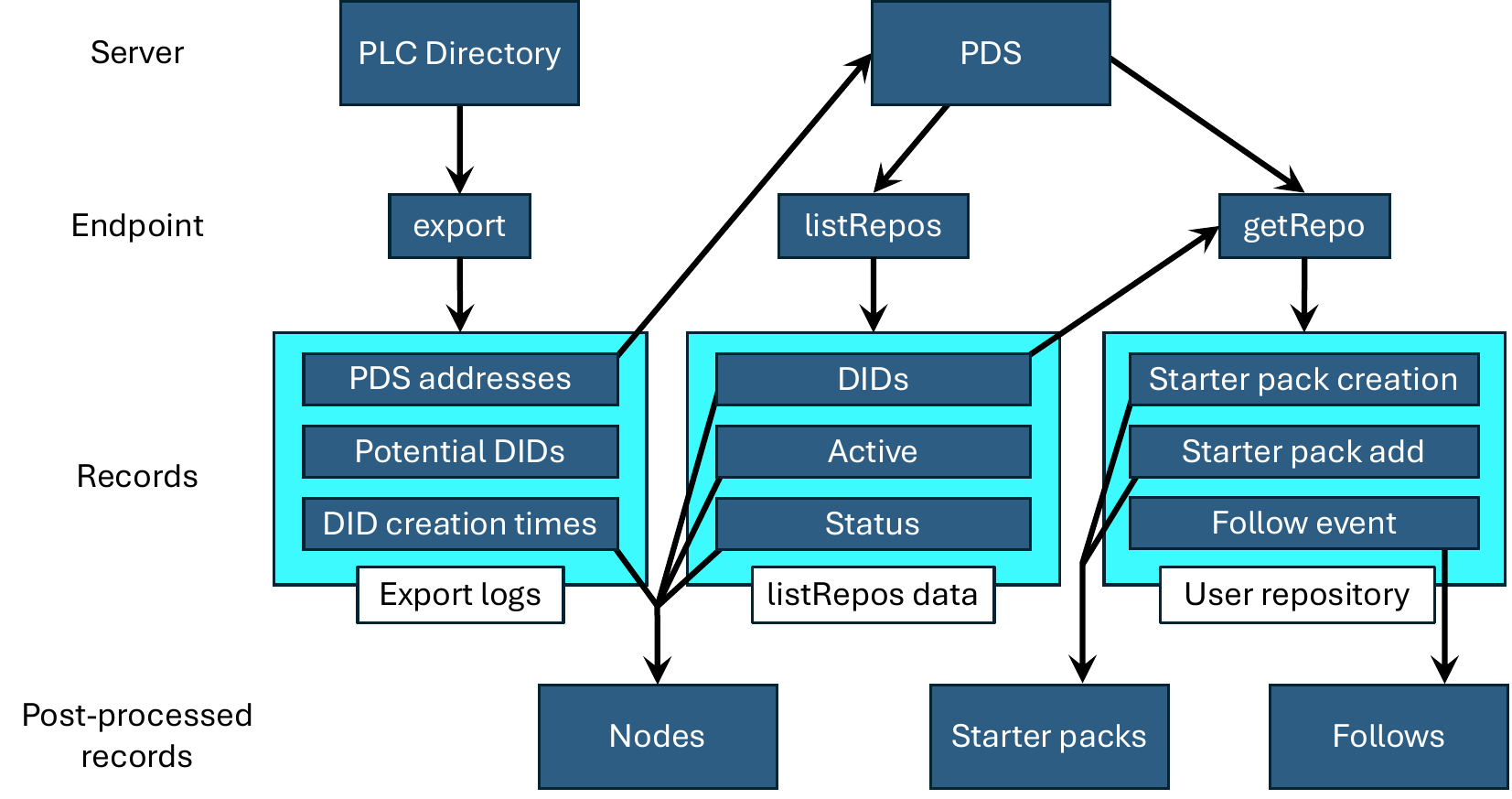}
    \caption{\textbf{Illustration of the data collection pipeline.} We export records from the Bluesky PLC Directory with the \texttt{export} endpoint to obtain the list of all possible PDSs and the list of possible DIDs with associated creation times. For each PDS in the list of PDSs obtained from the PLC Directory, we obtain the list of all DIDs hosted by that PDS along with their status with the \texttt{listRepos} endpoint. For active DIDs, we obtain the associated user repository containing all following and starter pack data using the \texttt{getRepo} endpoint. We then collate these three different records to obtain node, starter pack, and following datasets.}
    \label{fig:illustration}
\end{figure}

Collecting the data records was a multi-step process, requiring three steps for collecting the raw data from the Bluesky API as well as a final collation step. This process is illustrated in Fig.~\ref{fig:illustration}.

All user accounts on Bluesky have an associated decentralized identifier (DID).
DIDs provide a persistent global reference for all account information and all queries to the Bluesky API to collect user information such as the accounts followed or the list of starter packs require a DID as a query parameter.
DIDs allow us to link records in both the querying and collation steps.

\subsection*{Exporting records from the PLC Directory}

The complete list of DIDs and all operations performed on these DIDs, such as creation, modification, and deletion, are stored in Bluesky's "Public Ledger of Credentials (PLC)" hosted on the DID PLC Directory (\href{https://plc.directory}{https://plc.directory}), a resource that is provided and maintained by Bluesky itself.
A DID in the PLC directory, however, may not have an associated user account; as of Oct. 18, 2025, the PLC Directory contained 65,146,797 unique DIDs, in contrast to the 38,939,515 DIDs hosted on PDSs and the 39,650,447 accounts present in our dataset.
However, the PLC Directory is useful for two reasons: first, the first appearance of a DID in the directory corresponds with its creation time and second, the directory lists all of the personal data servers (PDSs) on which user account data is hosted.

The \texttt{export} endpoint returns a cursor which allowed us to page through in batches of 1,000 records and download the contents of the PLC Directory, which took roughly 24 hours.
We queried the PLC Directory for all entries with time stamps before October 18th, 2025 12:00:00 UTC.
From this log file, we collected the list of possible PDS instances and the list of all possible DIDs with associated creation times, with the knowledge that they may not all be valid user accounts.

\subsection*{Querying the Personal Data Servers}

While DIDs listed in the PLC Directory may not correspond to a Bluesky account, DIDs listed by a PDS are guaranteed to be active or inactive user accounts and we use this list of DIDs as our node list.
Our method for querying user IDs by getting the list of DIDs from the PDS instances is more efficient for two reasons: first, we do not consider the 25M+ DIDs which have no associated user account, and second, querying the list of all accounts in a PDS along with their activity status allows us to avoid querying inactive accounts.

Querying the PDSs is a two-step process; first, the list of active user repositories hosted by a particular PDS is obtained and second, these user repositories are downloaded.

\subsubsection*{Obtaining the list of users in each Personal Data Server}

While in principle one could collate the list of all DIDs by starting from a well-connected user account and performing a breadth-first search through the following relationships, the network is disconnected and, as seen in Fig.~\ref{fig:follows_network_stats}(c), some of the strongly-connected components are non-negligible in size.
For this reason, we chose to first compile the list of accounts from the accounts listed across all PDSs and collect all following relationships and starter packs from each account.

We populated an asynchronous queue with PDS addresses and attempted to query each PDS in this queue through the \texttt{com.atproto.sync.listRepos}.
Querying this endpoint returns a cursor which allowed us to page through in batches of 1000 DIDs and collect the entire list of accounts hosted by that PDS.
Upon an unsuccessful query, we retried three times with exponential back-off if the status was 408, 429, 500, or 502-504.
Of the 7,731 PDS addresses obtained from the PLC directory, only 3,163 were able to be queried, while 4,568 could not be queried.
While we cannot be absolutely sure that these repositories cannot be queried, we manually double-checked a number of them.
The output of the \texttt{listRepos} endpoint is in JSON format and the "repos" field contains a list of JSON objects, where each entry contains a DID, a boolean indicating whether the associated account is active, and, if not, the reason for its inactivity.
Accounts become inactive either because the platform suspended their account or the user deactivated their account.

We obtain each DID's creation time from the PLC directory as described above and if the DID was created prior to October 18th, 2025 12:00:00 UTC, we add the node to the list of nodes.
This list of nodes contained the DID, the creation time, whether it was active, and its activity status (We added a "normal" status if it was active).
If the DID was also active, we added the DID with its associated PDS address to another asynchronous queue which was used to query the user repositories.
Interestingly, several PDSs contained identical lists of user accounts and in this case, we added the first occurrence to the queue and ignored all other occurrences.

\subsubsection*{Obtaining user records}

We attempted to obtain the associated repository of each DID in the asynchronous queue of (DID, PDS) combinations generated in the previous step.
For each active DID,  we queried the \texttt{com.atproto.sync.getRepo} endpoint exposed by its host PDS.
As with the \texttt{listRepos} endpoint, upon an unsuccessful query, we retried three times with exponential back-off if the status was 408, 429, 500, or 502-504.
Of the 36,689,265 user repositories queried, 36,485,432 queries were successful and 203,833 queries were unsuccessful.
Unsuccessful queries could be due to a number of factors, including permission issues which we noticed on a number of accounts, as well as server errors at the time of querying.
Nonetheless, we note that the repositories unable to be queried comprise roughly 0.6\% of the total, indicating a robust collection pipeline.
This repository contains all user actions, such as posts, reposts, likes, blocks, follows, and list/starterpack creation.
However, the repository is stored in Content Addressable aRchive (CAR) format, a binary format for efficient data storage and processing.
We used a modified version of the \texttt{atmst} Python package to convert user records from this binary format to JSONL format.

We saved each user repository as a separate JSONL file, filtering on the creation time of the record; we only keep records created prior to October 18th, 2025 12:00:00 UTC. We queried the list of accounts for two weeks, ending on Nov 2, 2025.

\subsubsection*{Technical details}

Because HTTP requests are an I/O limited operation, the Python \href{https://docs.python.org/3/library/asyncio.html}{\texttt{asyncio}} and \href{https://docs.aiohttp.org/en/stable/index.html}{\texttt{aiohttp}} packages offer an efficient way to query the Bluesky API.
We used 1,024 asyncio tasks to consume both the PDS queue and the queue of (DID, PDS) combinations, and in accordance with industry wisdom, we limited the number of per-host concurrent connections to 64.
We implemented a rate limiter which extracted the rate limits for each PDS from the response headers and ensured that the number of requests to each PDS was under this limit.
We paired this rate limiter with retry logic, using an exponential back-off if the initial query was unsuccessful.

\subsection*{Collating records}

As described in Fig.~\ref{fig:illustration}, we combine three separate data sources to create the node, following, and starter pack datasets.

\subsubsection*{Collating the starter pack network}

Starter packs are stored in two types of user records: one describing starter pack-level metadata, such as the creator and creation time, and another listing the DIDs and the time at which they were added to the starter pack.
We iterated through each user repository searching for \texttt{app.bsky.graph.starterpack} or \texttt{app.bsky.graph.listitem} records.
From the former, we extract the creation date, the starter pack creator, and the list identifier associated with the starter pack members.
From the latter, we collate lists of potential starter pack members comprising user accounts and feeds by their list identifier.
We then pair list members with the starter packs by joining on the list identifier.
Note that entities in a starter pack are not necessarily users, they can be any entity that can be "followed" on Bluesky (e.g. a list or a feed).
Lists can only be followed, however, by circumventing the standard user interface through the API.

We parallelized the reading of each user repository across 64 processes and placed each starter pack stored as a Python dictionary in a queue to be consumed by a single-threaded file writer, since writes are not atomic on a distributed file system.
The file writing operation saved the starter packs as a JavaScript Object Notation Lines (JSONL) file, where each starter pack is a line in the file.
We used the XGI Python package~\cite{landry_xgi_2023} to convert the starter pack network to the HIF standard~\cite{coll_hif_2025} and a hyperedge list.

\subsubsection*{Collating the following network}

User following events are stored in \texttt{app.bsky.graph.follow} user records.
Each record contains the following account, the account followed, and the time at which the follow occurred.
We parallelized the reading of each user repository across 64 processes and placed the list of all follows as a comma-separated string where each line is a following event in a queue to be consumed by a single-threaded file writer.
The file writing operation saved the following dataset as a Comma Separated Variables (CSV) File, where each following event is a line in the file.

\subsubsection*{Collating the account list}

While in principle, the account list should be completely specified by the list of DIDs aggregated across all PDS instances, in practice, the following and starterpack datasets refer to DIDs not listed in the PDSs.
The most plausible explanation is that we unsuccessfully queried the hosting PDS of this DID; despite retrying failed queries, it is possible that there are active servers that we were not able to query.
We collate the list of all accounts by compiling lists of (1) all accounts listed in the following dataset, (2) all accounts listed in the starter pack dataset, and (3) all accounts listed by hosting PDSs and then taking the set union between these three lists.
Because there are accounts in the following network and accounts in the starter pack network which are not listed by hosting PDSs, 791,650 accounts do not have statuses.
For each DID in the collated list of DIDs, we start by attempting to pair this DID with a record from the listRepos record since this offers the richest data and if this record does not exist, we fall back to the creation time extracted from the PLC directory logs.

\subsection*{Anonymizing user data}

Although all user data is available via the public Bluesky API and users can require authentication to access their public user data from the API, we nonetheless take reasonable steps to prevent re-identification of users.
We recognize, however, that with enough effort, that it is possible to re-identify all users.
However, we assert that this effort will be comparable to querying the API from scratch.
We anonymize our dataset in three ways: (1) by converting the DIDs of all users and feeds to unique integer identifiers, (2) by removing the name and description of all starter packs, and (3) by coarsening the precision of all timestamps to one day.
We anonymize the accounts in the starter pack network, the following network, and the list of all accounts by mapping each DID to an integer.
We do the same for each starter pack URI, converting it to an integer.
Then we iterate through every starter pack and every following relationship using this mapping from DIDs and URIs to unique integers.
This process ensures that while users cannot be identified, node IDs are consistent across the account list, the following network, and the starter pack network.

\subsection*{Limitations}

The approach we utilized in collecting the starter pack and following data comes with limitations, however.
The number of active accounts, starter packs, and following relationships on the Bluesky platform is rapidly changing and any collection method not taking this into account will not perfectly capture the network structure.
Using the PLC Directory as a static directory does not consider accounts created after this list was collected.
This could potentially lead to missing accounts that have been migrated to a PDS created after the last entry of the PLC directory collected, but before the accounts were successfully queried.
Because we remove all events with creation dates after October 18th, 2025 12:00:00 UTC, we will not, in principle, collect network data for DIDs absent from the PLC directory; however, we are unable to collect accounts that are deactivated, starter packs that are deleted, and accounts that are unfollowed before they have been successfully queried.
Despite this temporal filtering, however, there are nonetheless a small number of accounts and events with timestamps later than this cutoff time.
We suspect that this could be a server-side issue or a problem with time zone conversion.
However, they are also associated with following events which fall within the specified cutoff time.
To this point, time stamps and user data can be arbitrarily set by users if they interact with Bluesky directly through the AT Protocol and not through the Bluesky application.
We discuss this limitation in the Technical Validation section.
It is also not strictly true that the network data we collect is in the PLC directory; there are 3,441 accounts in the list of accounts we collate which are not present in the PLC Directory.
One way around some of these limitations is by downloading streaming data from the Bluesky Firehose API while backfilling all events for all users, as is done in Ref.~\cite{balduf_bootstrapping_2025}.
The downside to this approach is that it is an immense amount of data; roughly 27GB of JSON Lines (JSONL) data per day at the time of writing.
We believe that our approach strikes the balance of being computationally manageable, while surveying close to the entire network.

As the Bluesky social media platform continues to grow and evolve, it is important to augment this dataset with new data.
Our method can be used to periodically create "snapshots" of the network, but this raises the important question of the timescales of these dumps.
Ref.~\cite{balduf_bootstrapping_2025} noted that over the course of their study, 20\% of starter packs were deleted.
Our static method cannot observe these deletions, nor the precise time at which they occur.
Long term, the most robust method for complete temporal data is through a pipeline using the Bluesky Firehose API, but it is beyond the scope of this dataset.

\subsection*{Ethical Considerations}
This data collection was ruled an exempt protocol by the Institutional Review Board (IRB) at Northeastern University (No. 25-01-22) and by the Institutional Review Board for the Social and Behavioral Sciences (IRB-SBS) at the University of Virginia (No. 7332).
All of the data obtained is publicly accessible (without authentication or a Bluesky account) via the Bluesky API.
Additionally, this study was granted a waiver of informed consent from both the IRB at Northeastern University and the IRB-SBS at the University of Virginia; arguably, obtaining informed consent for all users' inclusion in this dataset would constitute a larger risk to users' privacy than the collection of publicly available data.
Nonetheless, we recognize that this dataset could enable large-scale identification of accounts' following relationships.
We therefore redact all user DIDs (decentralized identifiers, used for long-term, persistent identification of entities in the AT Protocol that Bluesky uses) and truncate the timestamps on all starter pack and account creation events to include only the date; i.e., the level of precision at which timestamps are reported is one day.
The details of this process are described in the section detailing the anonymization of user data.

\section*{Data Records}

This study releases a list of anonymized accounts and two large-scale networks: (1) a list of starter packs on Bluesky and (2) a list of user following relationships.
All data is available on the Social Media Archive (SOMAR), which is a project by the Inter-university Consortium for Political and Social Research. 
The datasets described in this paper are available in the data collection titled "A Blue Start: A large-scale pairwise and higher-order social network dataset"~\cite{smith_blue_2026}.
The dataset is available under a CC-BY license with an additional clause that stipulates users will not attempt to reidentify individuals in the dataset.

\subsection*{List of nodes}
We provide a list of nodes in the network with accompanying metadata for each node.

\xhdr{JSON Lines (\texttt{deidentified\_nodes.jsonl.gz}, 221.4 MB).}
Each line in this JSON Lines file is a JSON dictionary with mandatory fields "node-id," representing the node's anonymized integer identifier, and "active," which is a Boolean attribute indicating whether the node was active or taken down at the time of data collection. The optional fields provide further information about the nodes: "date-created" is the date (in YYYY-mm-dd format) on which the node was created; \texttt{status} is one of "deactivated," "normal," or "takendown," indicating whether the node is inactive (initiated by the user), active, or inactive (initiated by the platform), respectively.

\xhdr{CSV (\texttt{deidentified\_nodes.csv.gz}, 187.7 MB).}
Each line in this compressed CSV file contains the fields "id," "date-created," "active," and "status." The "id" field represents the node's anonymized integer identifier; the "date-created" field is the date (in YYYY-mm-dd format) on which the node was created; the "active" field is a Boolean attribute indicating whether the node was active or taken down at the time of data collection; and the "status" field is one of "deactivated," "normal," or "takendown," indicating whether the node is inactive (initiated by the user), active, or inactive (initiated by the platform), respectively.

\subsection*{Bluesky starter packs}

We release the starter pack dataset in three different formats to ensure maximum cross-compatibility.

\xhdr{JSON Lines (\texttt{deidentified\_starterpacks.jsonl.gz}, 86.1 MB).}
We provide a JSON Lines file of all scraped starter packs' contents and metadata, intended for computational analysis. This file contains a list of starter packs; each starter pack contains the fields described in Table \ref{tab:json_file}. 

\begin{table}[h]
\centering
\begin{tabular}{m{3.5cm}m{2.5cm}m{8cm}}\toprule
Field &  Type & Description \\
\midrule
\multicolumn{3}{l}{\textbf{Top-level attributes}}\\
\midrule
\texttt{pack-id} & integer & Anonymous integer identifier of the starter pack list \\
\texttt{creator-id} & integer & Anonymous integer identifier of the starter pack’s creator \\
\texttt{date-created} & date & The date on which the starter pack was created \\
\texttt{members} & array & List of starter pack members\\[0.1in]
\multicolumn{3}{l}{\textbf{Entries in the \texttt{members} array}}\\
\midrule
\texttt{id} & integer & The member's anonymous integer identifier\\
\texttt{date-added} & date & The date on which the member account was added to the starter pack\\
\bottomrule
\end{tabular}
\caption{\label{tab:json_file} Fields for the JSON Blob and the HIF-compliant files of the starter pack network.}
\end{table}

\xhdr{Hypergraph Interchange Format (\texttt{deidentified\_starterpack\_hif.json.gz}, 111.5 MB).}
This file is a \href{https://github.com/pszufe/HIF-standard}{HIF-compliant} JSON file.
The HIF standard provides a cross-platform standard for sharing attributed higher-order network data.
The edge attributes ("id", "creator-id", and "date-created") are labeled as described in Table \ref{tab:json_file}. Each entry in the "incidences" field has attributes "edge," corresponding to a starter pack (edge) ID, "node," corresponding to a node ID, and "date-added," indicating when the node was added to the starter pack, as explained in Table \ref{tab:json_file}. The attributes included in the "nodes" field are "node" (node ID), "active," and "status" as seen in the node list described previously.

\xhdr{Hyperedge list (\texttt{deidentified\_starterpack\_edgelist.csv.gz}, 46.6 MB).}
We also provide a hyperedge list stored as a gzipped CSV file. Each line in this file corresponds to a starter pack, and each entry in the line is a member of the starter pack. Each line contains entries separated by commas; a single line corresponds to a single starter pack. For example, the line "1,2,3,4" indicates that users with anonymous integer identifiers 1, 2, 3, and 4 belong to the same starter pack.

\subsection*{Bluesky following network}

We also release the following network in two different formats.

\xhdr{Edgelist CSV (\texttt{deidentified\_follows\_edgelist.csv.gz}, 14.9 GB).}
We provide the following network dataset as a compressed CSV file.
Each line $(i, j, d)$ in the file indicates that user $i$ followed user $j$ on date $d$, where $d$ is provided in YYYY-mm-dd format. Here, $i$ and $j$ are anonymous integer identifiers that map to user DIDs.
As described in the section detailing the anonymization of user data, these anonymous integer IDs can be linked to creators and members of starter packs.

\xhdr{Edgelist Parquet (\texttt{deidentified\_follows\_edgelist.parquet}, 12.6 GB).}
This data artifact has the same format as the CSV, but it is encoded in the more efficient Parquet format and has column names \texttt{from}, \texttt{to}, and \texttt{date\_followed}.

\section*{Technical Validation}

Here, we describe the statistical properties of both the starter pack network and the following network, demonstrating their validity and utility for researchers analyzing social network data.

\subsection*{Key Events}
As we studied the temporal elements of the datasets (i.e. node creation, starter pack creation, following tie formation), we noticed spikes in activity that appear to be driven by key events; for example, a spike in node creation, starter pack creation, and following tie formation occurred when X/Twitter was banned in Brazil on August 30, 2024. We therefore include a table of these events; each event is indicated in the temporal plots, where relevant, using an integer (indicating an event that was important to multiple datasets) or a single letter (indicating an event pertinent only to the starter pack dataset) to identify it. Table~\ref{tab:dates} is color-coded to distinguish on-platform events from off-platform ones.

\begin{table}[ht]
\centering

\begin{tabular}{ c  l  l }
\toprule
\textbf{Identifier} & \textbf{Event} & \textbf{Date} \\
\midrule
\rowcolor{color2!17}1 & Beta version of Bluesky opens; invitation codes are needed to join. & February 17, 2023 \\
\rowcolor{color2!17}2 & Bluesky is open to the general public & February 6, 2024 \\
\rowcolor{color4!17}3 & X blocks popular accounts in Brazil & April 6, 2024 \\
\rowcolor{color4!17}4 & X formally allows adult and graphic content & June 2, 2024 \\
\rowcolor{color2!25} A & Bluesky launches starter pack feature & June 26, 2024 \\
\rowcolor{color4!17}5 & X is banned in Brazil & August 30, 2024 \\
\rowcolor{color4!17}6 & New X terms of service allow blocked users to see posts & October 16, 2024 \\
\rowcolor{color4!17}7 & New terms of service allow X to train AI on user content & November 15, 2024 \\
\rowcolor{color4!17}8 & Donald Trump sworn in as U.S. president & January 20, 2025 \\
\rowcolor{color4!17}9 & Nationwide Hands Off protests in U.S. & April 5, 2025 \\
\bottomrule
\end{tabular}
\caption{\label{tab:dates} Significant events in the history of the Bluesky social media platform.}
\end{table}

\begin{table}[ht]
\centering
\begin{tabular}{lr}
\toprule
\textbf{Nodes} & \textbf{Count} \\
\midrule
\rowcolor{color2!17}Total users & 39,650,447 \\
\rowcolor{color4!17}Extant users & 37,479,031 \\
\rowcolor{color2!17}Users from listRepos & 38,858,797 \\
\rowcolor{color4!17}Active users & 36,687,381 \\ 
\rowcolor{color2!17} Inactive users & 2,171,416 \\
\rowcolor{color2!7} Taken down (moderation) & 1,808,660 \\
\rowcolor{color2!7} Deactivated & 362,756 \\ 
\rowcolor{color4!17}Users with no status & 791,650 \\
\bottomrule
\end{tabular}
\caption{\label{tab:nodes} User statistics.}
\end{table}

\subsection*{Nodes}
\begin{figure}
    \centering
    \includegraphics[width=0.75\linewidth]{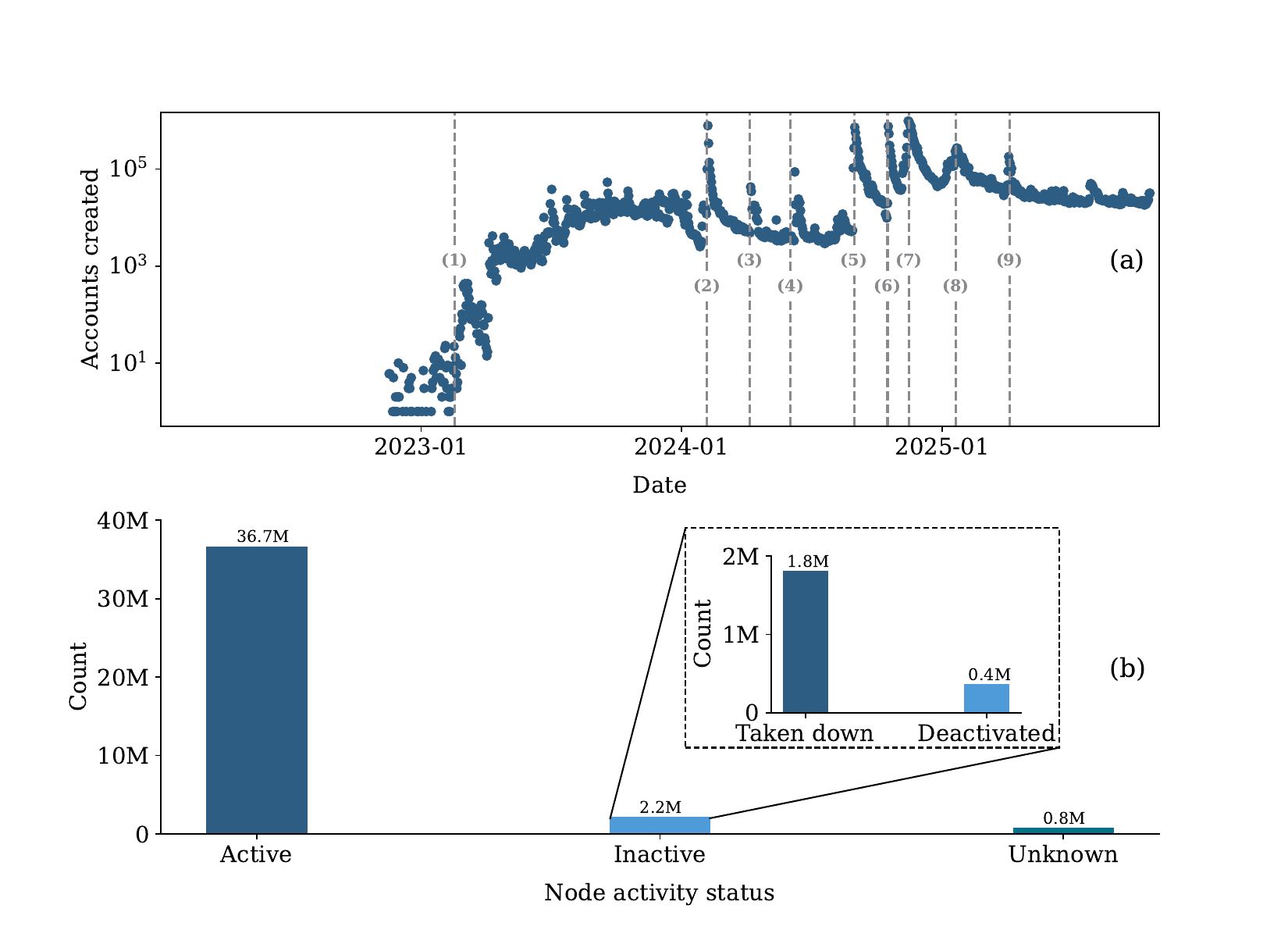}
    \caption{\textbf{Bluesky account creation and activity.} Panel (a) plots the volume of account creations per day, overlaid with events significant to Bluesky's history. Panel (b) plots the activity statuses for all accounts, with inactive accounts broken out into accounts taken down through moderation and deactivation.}
    \label{fig:node_creation}
\end{figure}

As seen in Fig.~\ref{fig:node_creation} and Table~\ref{tab:nodes}, the node dataset consists of 39.7 million accounts, of which 36.7 million are active, 2.17 million inactive, and 791k lacking an obtainable active/inactive status. Of the inactive accounts, 1.81 million had been taken down through moderation, and 363k had been deactivated. 

\subsection*{Starter pack data}

\begin{table}[ht]
\centering
\begin{tabular}{lr}
\toprule
\textbf{Starter Pack Network Statistic} & \textbf{Value} \\
\midrule
\rowcolor{color2!17}Nodes & 2,003,536 \\
\rowcolor{color2!17}Edges & 365,842 \\
\rule{0pt}{12pt}\textbf{Components}&\\
\midrule
\rowcolor{color4!17}Number & 409 \\
\rowcolor{color4!17}Largest & 1,997,488 (99.7\ of starter pack nodes) \\
\rule{0pt}{12pt}\textbf{Starterpacks created} & \\
\midrule
\rowcolor{color2!17}Min / Max & 1 / 252 \\
\rowcolor{color2!17}Mean / Median & 1.162 / 1 \\
\rowcolor{color2!17}Mode & 1 \\
\rowcolor{color2!17}Users with none & 84.3\% \\
\rule{0pt}{12pt}\textbf{Starterpack Size} & \\
\midrule
\rowcolor{color4!17}Min / Max & 0 / 4,069 \\
\rowcolor{color4!17}Mean / Median & 34.724 / 24 \\
\rowcolor{color4!17}Mode & 50 \\[0.2cm]
\rule{0pt}{12pt}\textbf{Degree} & \\
\midrule
\rowcolor{color2!17}Min / Max & 1 / 175,159 \\
\rowcolor{color2!17}Mean / Median & 6.341 / 1 \\
\rowcolor{color2!17}Mode & 1 \\
\rowcolor{color2!17}Top-5 account frequencies & 47.88, 13.95, 11.59, 10.9, 10.47\% \\
\rule{0pt}{12pt}\textbf{Line Graph} & \\
\midrule
\rowcolor{color4!17}Density for $s=1\dots 5$ & 0.2923, 0.0665, 0.034, 0.0207, 0.0139 \\

\rule{0pt}{12pt}\textbf{k-core} & \\
\midrule
\rowcolor{color2!17}k-core ($k \ge 1000$) size & 772 \\
\rule{0pt}{12pt}\textbf{Communities} & \\
\midrule
\rowcolor{color4!17}Number & 503 \\
\rowcolor{color4!17}Top-5 community sizes & 434,207; 333,148; 156,110; 126,567; and 98,434\\
\rowcolor{color4!17}Mean normalized edge entropy & 0.16 \\
\bottomrule
\end{tabular}
\caption{\label{tab:sp_stats}Starter pack network statistics}
\end{table}
The starter pack data comprises 2,003,536 unique user and feed IDs and 365,842 starter packs.
These statistics corroborate the findings of Ref.~\cite{balduf_bootstrapping_2025} which assembled a dataset of 335,416 starter packs.
The discrepancy in sizes between these two datasets are attributable to the time of collection, the method of collecting the dataset, and the deletion of existing starter packs.
Ref.~\cite{balduf_bootstrapping_2025} collected all of the starter packs from their introduction up to, and not including, January 1st, 2025, whereas with our method, we collected the list of DIDs on October 18, 2025 and queried the list of starter packs over several days ending on November 2, 2025.
Out method relied on querying the Bluesky Sync API, whereas Ref.~\cite{balduf_bootstrapping_2025} used the Bluesky Firehose API and the Bluesky Sync API (For more details, see the section detailing limitations of our methodology) to collect all starter packs.
The advantage of the approach in Ref.~\cite{balduf_bootstrapping_2025} is that it can collect starter packs even if they are later deleted; it is estimated that 20\% of all starter packs were deleted by January 1st, 2025~\cite{balduf_bootstrapping_2025}.

Our analysis of the starter pack data is from a higher-order network perspective; particularly, we model starter packs as higher-order interactions, or \textit{hyperedges}, and the collection of starter packs as a higher-order network, or \textit{hypergraph}.
Higher-order network analysis can uncover network structure across different scales by modeling hyperedge inclusion~\cite{larock_encapsulation_2023,landry_simpliciality_2024}, overlap~\cite{aksoy_hypernetwork_2020}, and interaction size~\cite{landry_filtering_2024}.
Because starter packs are not dyadic, higher-order network analysis represents a natural framework to study this system.
We expect this analysis to yield important insights into the group overlap structure and, more importantly, how groups and and pairwise ties can co-evolve.

\begin{figure}[H]
    \centering
    \includegraphics[width=0.75\linewidth]{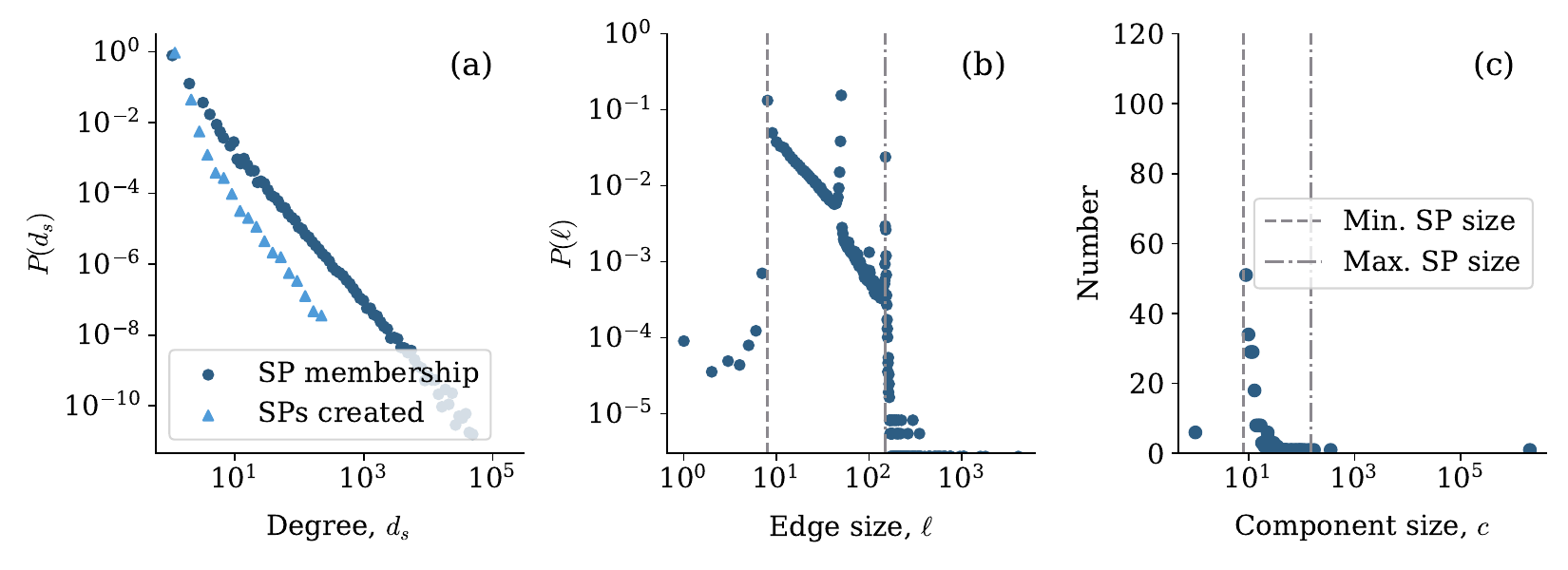}
    \caption{\textbf{Basic higher-order starter pack statistics.} Panel (a) plots the probability distributions of both the number of starter packs to which a user account belongs and the number of starter packs that each user account has created. Panel (b) plots the starter pack size distribution, with the minimum and maximum starter pack sizes illustrated. Panel (c) plots the sizes of the connected components with the minimum and maximum starter pack sizes illustrated.}
    \label{fig:sp_stats}
\end{figure}
We begin our analysis by examining the distribution of the number of starter packs that users belong to and create, as seen in Fig.~\ref{fig:sp_stats}(a) and Table~\ref{tab:sp_stats}.
Although the number of starter packs that users create varies widely, it is also quite homogeneous; while the majority (84.3\%) of accounts in the network did not create a starter pack, those who did created 1.16 starter packs on average.
Fig.~\ref{fig:sp_stats} shows the hyperedge (starter pack) size distribution in Fig.~\ref{fig:sp_stats}(b) which, interestingly, is trimodal, with peaks at 8, 50, and 150.
The peaks at 8 and 150 correspond to the minimum and maximum allowed starterpack sizes, and the peak at 50 is due to the Bluesky starter pack feature called "Make one for me," which pre-populates a starter pack with up to 50 accounts and feeds that a user follows.
The distribution of connected component sizes in Fig.~\ref{fig:sp_stats}(c) shows a giant connected component of 1.997M users accompanied by 408 components with all but two smaller than 1,000 users, the vast majority (383) of which are isolated starter packs.

\begin{figure}[H]
    \centering
    \includegraphics[width=0.75\linewidth]{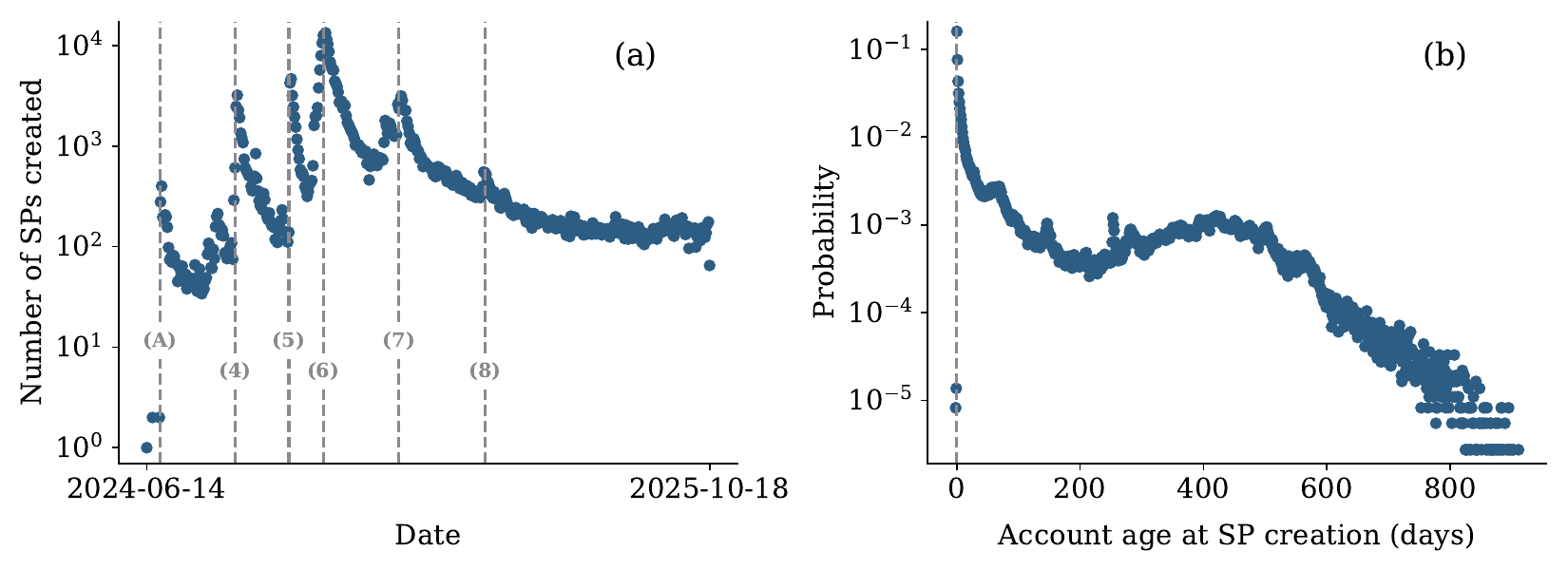}
    \caption{\label{fig:temporal_sp_stats}\textbf{Temporal starter pack statistics.} Panel (a) plots the number of starter packs created each day and panel (b) plots the distribution of time elapsed between when a Bluesky account was created and when that account created a starter pack. The vertical lines in panel (a) correspond to relevant events enumerated in Table~\ref{tab:dates}}
\end{figure}
Analyzing the times at which starter packs were created, we see six sharp peaks characterized by discontinuous jumps in the number of starter packs created and exponential decay after peak activity as illustrated in Fig.~\ref{fig:temporal_sp_stats}(a).
These peaks correspond to notable events in the history of Bluesky and X, as well as relevant U.S. political developments, which are detailed in Table~\ref{tab:dates}.
Fig.~\ref{fig:temporal_sp_stats}(b) plots the distribution of time between when a user makes a Bluesky account and when they create a starter pack. 
This plot exhibits two interesting features: first, a few starter packs were created by accounts with negative ages, and second, most starter packs are made by very recently created accounts. 
Negative account ages are physically impossible; starter packs cannot be created prior to the creation of the associated user account.
However, users have the ability to change their user data arbitrarily through the Bluesky API and this can cause data artifacts such as these. 
We emphasize that the number of starter packs with erroneous time stamps is small; there are 8 starter packs with creation times prior to the creator's account creation time, or 0.002\% of all starter packs.

\begin{figure}[H]
\centering
\includegraphics[width=0.75\linewidth]{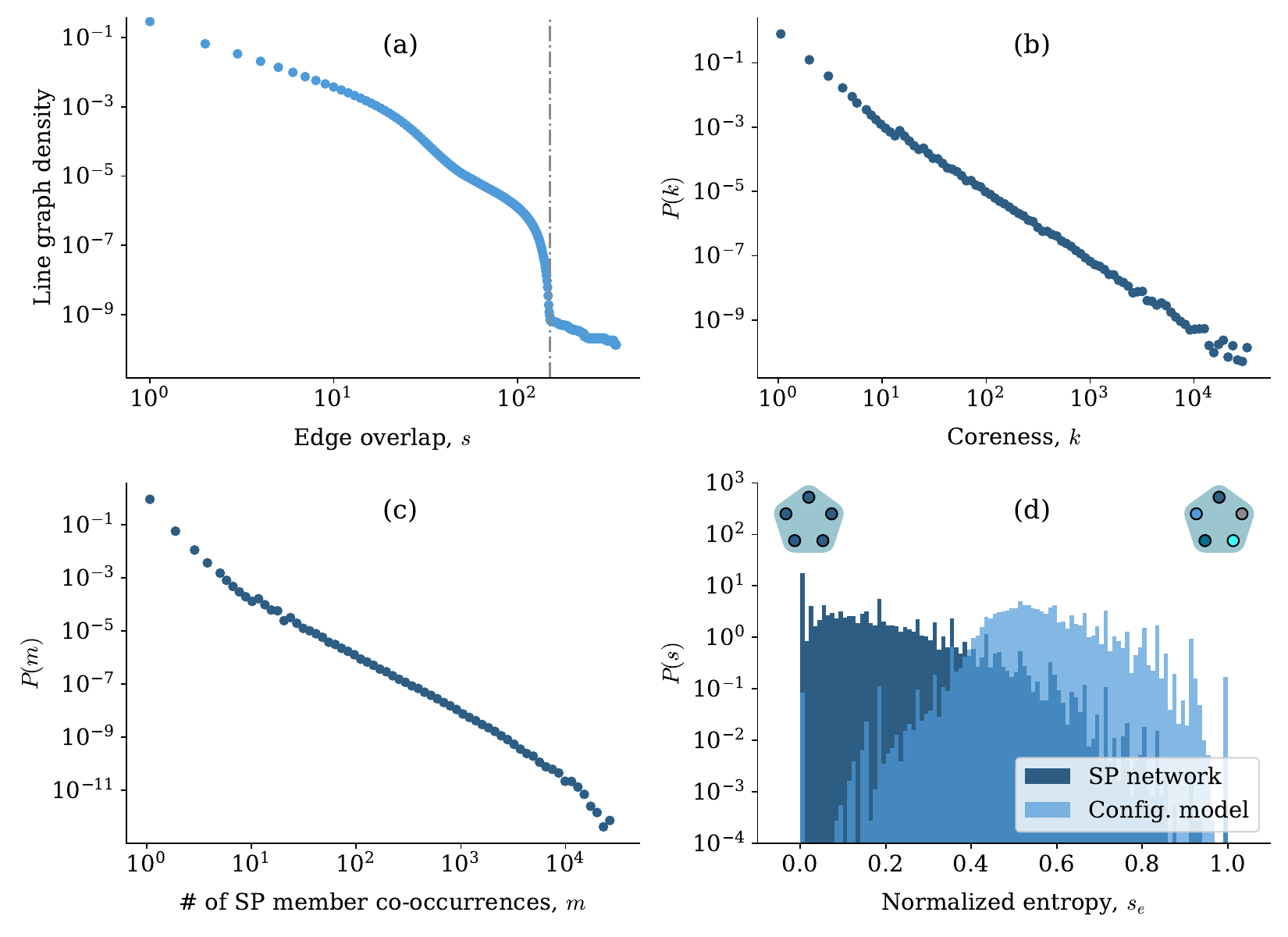}
\caption{\label{fig:mesoscale}\textbf{Mesoscale starter pack statistics.} Panel (a) plots the number of edges in the $s$-line graph of the starter pack network for all non-trivial values of $s$ (the dashed line denotes the maximum starter pack size of 150), Panel (b) plots the distribution of $k$-coreness, Panel (c) plots the probability of two user accounts appearing in $m$ different starter packs together, and Panel (d) plots the distribution of normalized entropy calculated for all starter packs. The illustrations of starter packs indicate at $s=0$, a starter pack completely contained in an inferred community and at $s=1$, a starter pack where every member belongs to a different community.}
\end{figure}

Line graph statistics are a powerful tool for analyzing higher-order network analysis and we use them to describe the edge overlap patterns in the starter pack network.
A line graph of a higher-order network is a network with nodes corresponding to hyperedges and edges connecting two nodes if the two hyperedges that they represent share at least one node in common.
Ref.~\cite{aksoy_hypernetwork_2020} generalized this idea by introducing $s$-line graphs which only connect two nodes with an edge if the hyperedges they represent share at least $s$ nodes in common.
We vary $s$, the edge overlap size, to get an idea of the strength of overlap across different regions of the network.
This can be quite computationally intensive and it is often necessary to exploit algorithms such as those described in Ref.~\cite{liu_highorder_2022}.
Because our analysis was confined to measuring the number of edges in the $s$-line graph, we were able to utilize a much more memory-efficient algorithm which focused on first finding cliques in the line graph and then looking for connections between edges in different cliques.
This approach allowed us to avoid storing all the edges in a clique and simply store the node labels of the clique, which reduced the memory requirements considerably.
Fig.~\ref{fig:mesoscale}(a) plots the number of edges in each $s$-line graph.
Note that these graphs are extremely large and dense; for $s=1$, 29\% of all possible edges are present in the line graph.
This is because several accounts are included in many starterpacks.
For example, the \textbf{@bsky.app} account is involved in 48\% of all starter packs, and the top-5 most included accounts are all in greater than 10\% of starter packs.
This higher-order network is far more dense and overlapping than many other existing higher-order datasets, providing an exciting opportunity for studies on higher-order network formation and higher-order algorithm development.

We perform a hypergraph $k$-core decomposition~\cite{amburg_planted_2021} and, as seen in Fig.~\ref{fig:mesoscale}(b), the distribution has a very heavy tail.
There are over 700 users with $k$-cores of over 1000 which is indicative of many nodes deep in the core of the hypergraph, implying the existence of large and intricate intersections between starter packs.
Fig.~\ref{fig:mesoscale}(c) plots the co-occurrence distribution of user pairs, i.e., the probability of two user accounts appearing in $m$ different starter packs together.
This can also be thought of as examining the distribution of weights in a weighted clique expansion of the hypergraph.
From ocular regression, the distribution appears piecewise linear, suggesting three distinct regimes.
There is a substantially high chance of having very large co-occurrence numbers.
These insights suggest complex co-occurrence patterns, with a plethora of users participating in a multitude of starter packs simultaneously.
In future studies, it will be fruitful to examine the co-occurrence of groups of $k$ nodes across different starter packs along the lines of techniques presented in Ref.~\cite{yoon_how_2020}.

To measure the community structure of the starter pack network, we run the Leiden~\cite{traag_louvain_2019} algorithm for two iterations on the unweighted pairwise projection of the largest connected component of the hypergraph to infer the community labels of all the nodes in the starter pack network, resulting in 503 clusters.
The largest connected component was used so that isolated edges and small components would not inflate the number of clusters calculated.
We use the inferred community labels to measure the Shannon entropy of each edge, normalized by the highest entropy configuration, where each node is in a different community. By this measure, the lowest entropy configuration ($s_e=0$) corresponds to a starter pack comprising nodes all belonging to the same community and the highest entropy configuration ($s_e=1$) corresponds to a starter pack comprising nodes all belonging to different communities.
Fig.~\ref{fig:mesoscale}(d) indicates that the starter pack network contains strong community structure when compared to the distribution of normalized edge entropy values computed for a configuration model, with means of 0.16 and 0.576 respectively.
These values are evidence of the correctness of the data; it is highly likely that the entropy values are much lower for an empirical dataset than for a random null model.
This analysis also indicates that the network is likely suitable as a benchmark for community detection algorithms.
In addition, even stronger community structure may also emerge if the network is filtered~\cite{landry_filtering_2024} to only include starter packs under a certain size or the network composed of starter packs with the most popular accounts removed.

\subsection*{Following data}

\begin{table}[ht]
\centering
\begin{tabular}{lr}
\toprule
\textbf{Following Network Statistic} & \textbf{Value} \\
\midrule
\rowcolor{color2!17} Nodes in network & 36,447,725 \\
\rowcolor{color2!17} Links in network & 2,416,311,437 \\
\rule{0pt}{12pt}\textbf{Largest connected component size} & \\
\midrule
\rowcolor{color4!17} Strongly connected & 20,495,220 (56.2\% of following nodes) \\
\rowcolor{color4!17}Weakly connected & 36,433,172 (99.96\% of following nodes) \\
\rule{0pt}{12pt}\textbf{In-degree} & \\
\midrule
\rowcolor{color2!17} Min / Max & 1 / 28,062,787 \\
\rowcolor{color2!17}Mean & 105.8 \\
\rowcolor{color2!17}Mean over all following nodes & 66.3 \\
\rowcolor{color2!17}Following nodes with in-degree 0 & 13,608,819 \\
\rule{0pt}{12pt}\textbf{Out-degree} & \\
\midrule
\rowcolor{color4!17}Min / Max & 1 / 844,408 \\
\rowcolor{color4!17}Mean & 70.38 \\
\rowcolor{color4!17}Mean over all following nodes & 66.3 \\
\rowcolor{color4!17}Followed nodes with out-degree 0 & 2,113,664 \\
\bottomrule
\end{tabular}
\caption{\label{tab:foll_stats}Following network statistics}
\end{table}
As seen in Table~\ref{tab:foll_stats}, the following network consists of 36.4 million distinct users and 2.42 billion directed following relationships linking a follower and a followed account.
The average out-degree for nodes following at least one account is 70.38, and the average in-degree for nodes that have at least one follower is 105.80.
The discrepancies in the mean degrees can be attributed to the fact that while 13,608,819 accounts in the following network have no followers, only 2,113,664 accounts in the following network do not follow any accounts.
When including degree-zero nodes, the mean in- and out-degrees are both 66.3.

\begin{figure}
    \centering
    \includegraphics[width=0.75\linewidth]{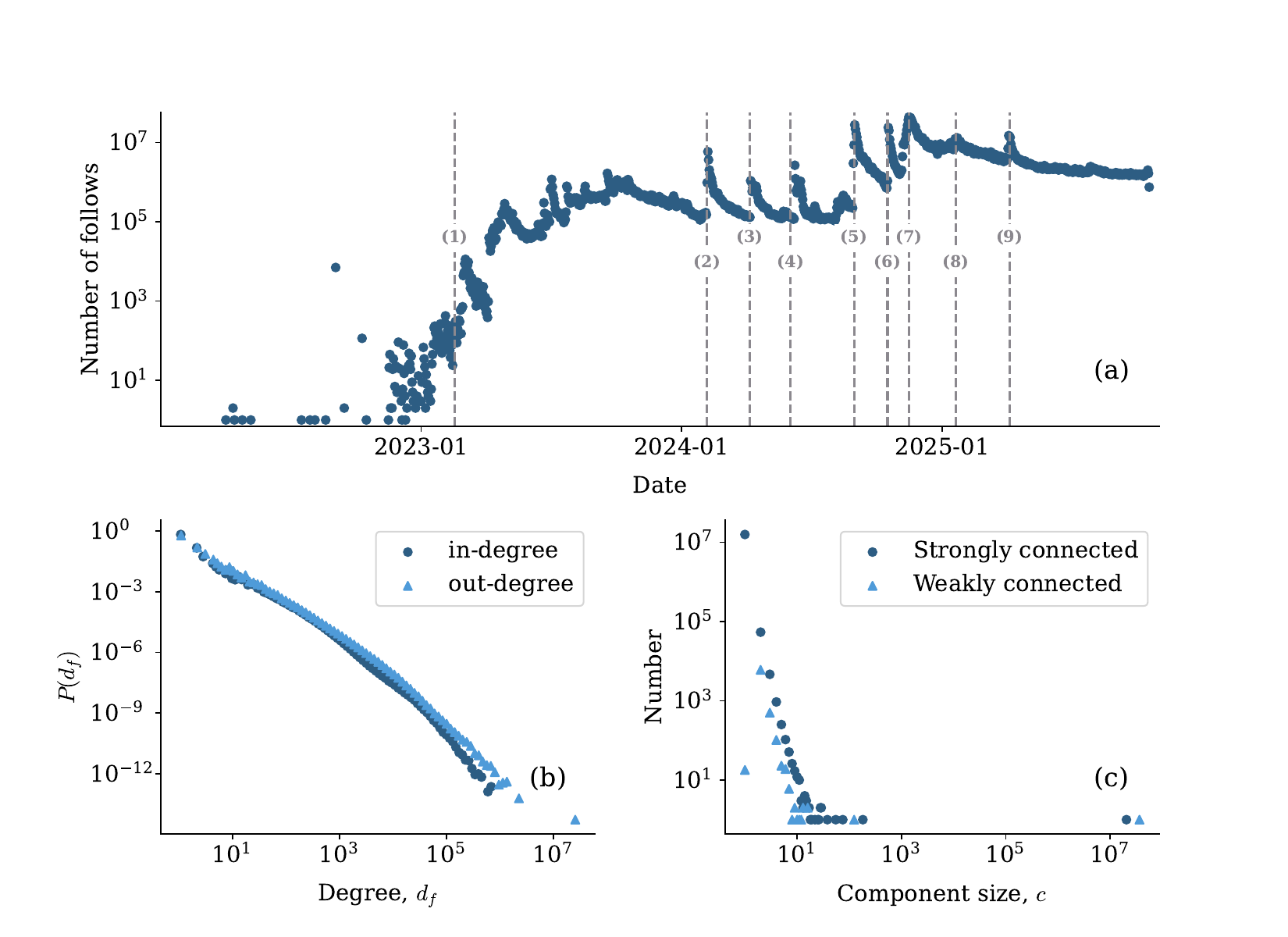}
    \caption{\label{fig:follows_network_stats}\textbf{Basic following network statistics.} Panel (a) plots the volume of following events for each day in our dataset, panel (b) plots the in-degree and out-degree distributions, and panel (c) plots the frequency of strongly and weakly connected component sizes. The vertical lines in panel (a) correspond to dates listed in Table~\ref{tab:dates}}
\end{figure}
When plotting the volume of new following relationships over time in Fig.~\ref{fig:follows_network_stats}(a), we have omitted data with erroneous time stamps by filtering out following events with time stamps from before Bluesky was established as a platform.
There are very few of these errors, however; 1 total follow has a verifiably incorrect time stamp.
We also check for impossible timestamps where the date on which the follow event occurred is strictly less than the date on which either the follower or the followed account was created. Again, these errors are rare; 147,655 follow events fit this criterion, or 0.006\% of follow events. As we discussed in the previous subsection, these impossible timestamps are likely a result of the fact that users can arbitrarily change their user data through the Bluesky API.
The time series of number of follow events is characterized by exponential growth in the first six months followed by more consistent daily follow volume with intermittent spikes due to notable world and platform events.
This approximate exponential growth can be attributed to the invite-only beta version of the app which allowed active users one new user per week with an invite code.
As the daily volume of the new follows slowed down, several notable events characterized large changes in follow volume.
For example, as illustrated in Fig.~\ref{fig:follows_network_stats}(a), (2) the public release of Bluesky on February 6, 2024~\cite{oremus_bluesky_2024} and (3) the announcement of the blocking of certain popular accounts from Brazil~\cite{weatherbed_elon_2024} on April 5, 2024 sparked a large increase in new follows on the Bluesky platform.
Similar to Fig.~\ref{fig:temporal_sp_stats}(a), we see the effect of several external events on the following behavior of Bluesky users, including changes in X's terms of service and political events in the United States.

We continue our analysis by looking at the in- and out- degree distributions; both distributions are highly heterogeneous and heavy-tailed as seen in Fig.~\ref{fig:follows_network_stats}(b).
In addition, we see that the in-degree distribution has a heavier tail than the out-degree distribution; not only are there more accounts following a large number of accounts, but the maximum number of accounts followed is larger than the number of followers of the most popular account.
Fig.~\ref{fig:follows_network_stats}(c) plots the distributions of both the strongly and weakly connected components in the following network.
The distribution of strongly connected component sizes contains one giant component of 20,495,220 nodes (or 56.2\% of the network), along with several strongly connected components of 100 nodes or fewer.
Looking at the distribution of weakly connected component sizes, we see an increase in the size of the giant component to 36,443,172 users (or 99.96\% of the network) and a decrease in the number of small connected components, indicating that while the vast majority of the following graphs is strongly reachable, there are still "in-components" and "out-components" up to several hundred nodes in size.
These components may or may not be reachable via snowball sampling~\cite{goodman_snowball_1961} depending on the starting node of choice.
A small number of components are truly disconnected from all other components via follower/following relationships.

Historically, the \texttt{created-at} field returned by the static Bluesky API corresponded to the time at which the followed account was created.
To verify that the time stamps of all following events truly correspond to following events, we calculate the standard deviation of the following times.
A standard deviation of 0 for the in- and out-degree corresponds to follows all occurring on the same day for followed and following accounts, respectively.
\begin{figure}[h!]
    \centering
    \includegraphics[width=0.75\linewidth]{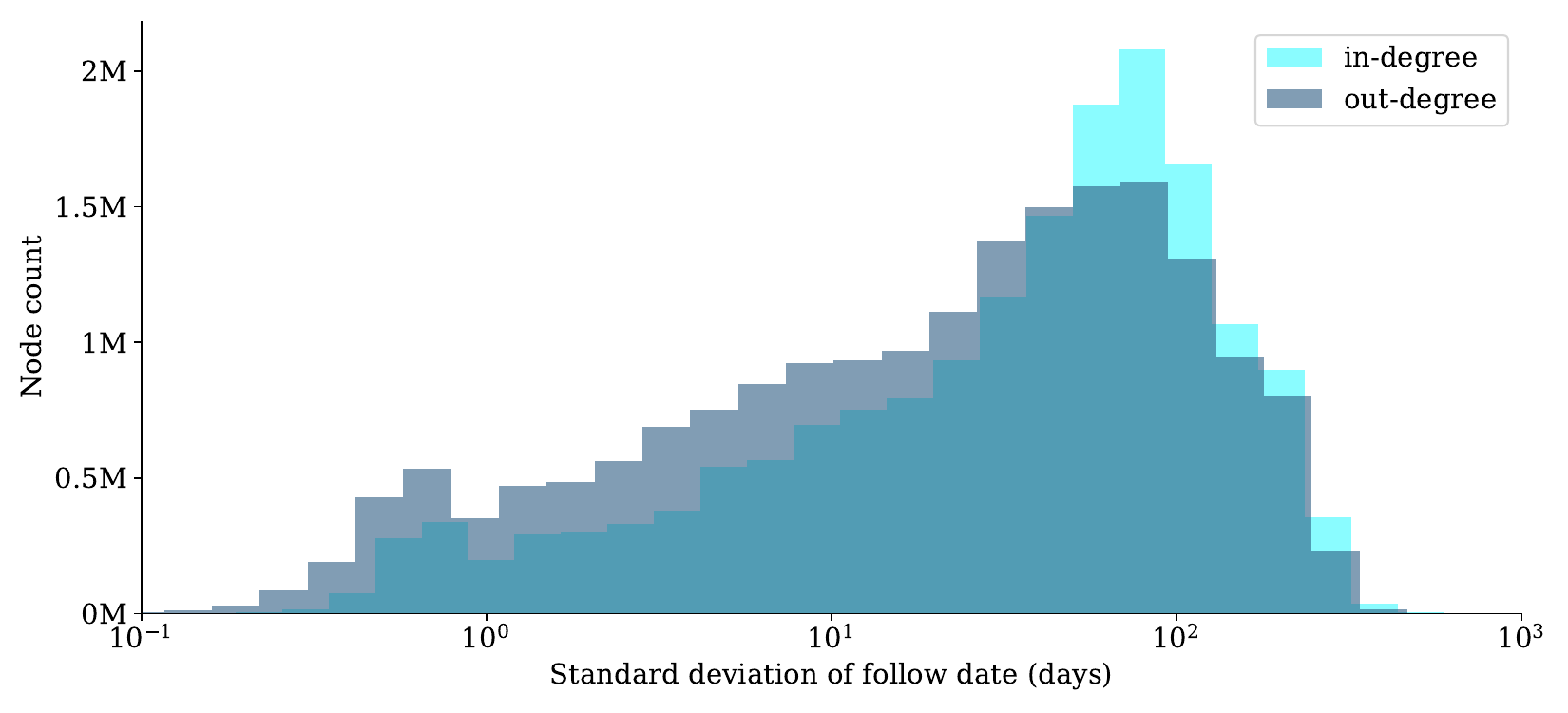}
    \caption{Distribution of standard deviations of follow dates.}
    \label{fig:degree_std}
\end{figure}
In Fig.~\ref{fig:degree_std}, we see that the standard deviations are not only non-zero, but they differ as well.
The distributions of these times make physical sense as well; it is more likely that a user follows most accounts within a small time range than for an account to be followed by many users within a small time range.

\subsection*{Comparing the starter pack and following network data}

\begin{figure}[H]
\centering
\includegraphics[width=0.75\linewidth]{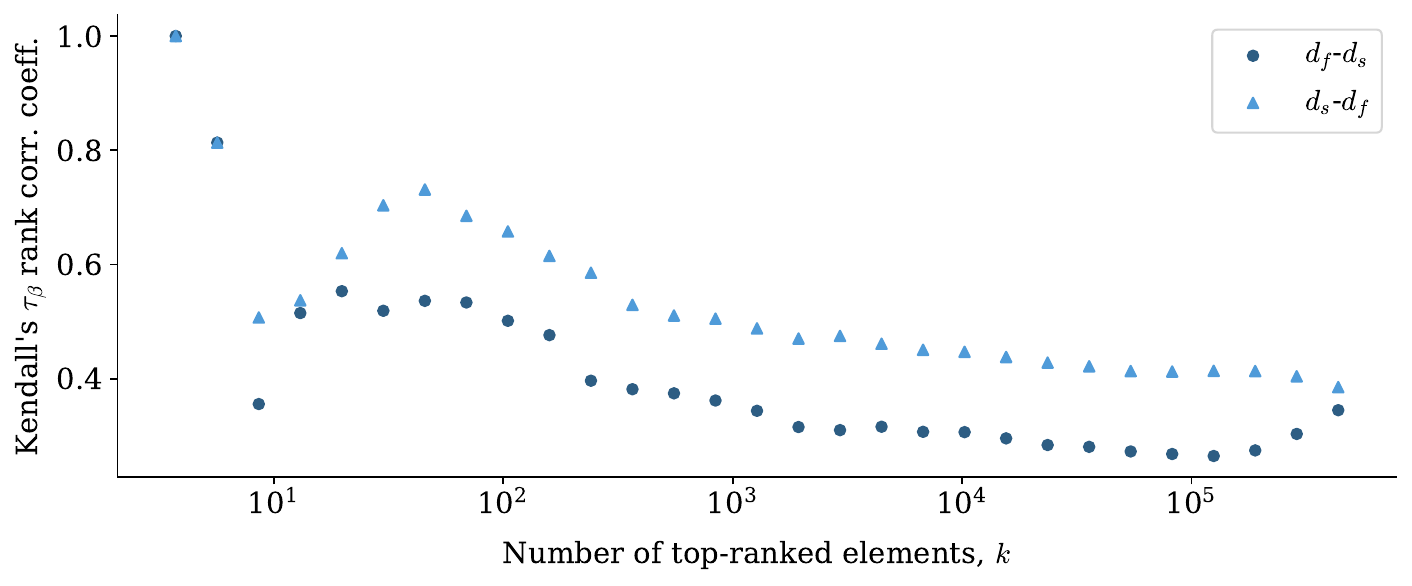}
\caption{\label{fig:hyper_graph_corr}\textbf{Comparing pairwise and higher-order network structure.} Kendall's $\tau_\beta$ rank correlation coefficient between degree centrality in the starter pack data ($d_s$) and the following data ($d_f$) as a function of number of top-ranked nodes compared.}
\end{figure}

The "A Blue Start" dataset provides a bridge between pairwise and higher-order network representations.
In Fig.~\ref{fig:hyper_graph_corr}, we compare the rankings induced by degree centrality in the starter pack hypergraph and the follows network.
In the starter pack hypergraph, the degree centrality $d_s$ of a node (user) is the number of hyperedges (starter packs) of which the user is a member, while in the following network, the degree centrality $d_f$ of a node (user) is the sum of its following and followed by counts.
Kendall's $\tau_\beta$ rank correlation coefficient between two rankings ranges from -1 (total disagreement) to 1 (total agreement) and systematically handles ties in ranking.
We start by computing $d_s$ and $d_f$ and only analyzing nodes appearing in both networks.
We 1) rank the top $k$ nodes according to $d_s$ and compute the correlation between the ranks of these same nodes according to $d_f$, presenting the results as $d_s-d_f$; and 2) rank the top $k$ nodes according to $d_f$ and compute the correlation between the ranks of these same nodes according to $d_s$ and present the results as $d_f-d_s$.
The fact that the maximal value of 1 is achieved at $k=1$ is not surprising, as the @bsky.app account is most central in both networks.
However, the rankings quickly diverge, with a weaker correlation of approximately 0.5 for $k=10$ ($d_s-d_f$); this association increases to around 0.7 at $k=100$ and decreases with increasing $k$.
Overall, relatively low correlation among the importance assigned to nodes in the higher-order (starter pack) and dyadic (following network) data demonstrates that the two contain complementary information about user importance, and that importance is conditioned on the context in which user interactions are measured.

\section*{Usage Notes}
Processing the following network can be computationally intensive; The CSV file contains 2.4B lines of following relationships and is 71 GB when inflated.
Simply loading this dataset using the \texttt{pandas} Python package~\cite{mckinney_data_2010,pandasdevelopmentteam_pandasdev_2020} requires over 200 GB of memory.
The size limitation of a \texttt{pandas} dataframe tends to be the amount of available memory and we encourage users of this dataset to verify that they have enough physical resources to process the following network data.
To parallelize the processing of these data, we recommend loading the gzipped CSV or Parquet files with \href{https://pola.rs/}{\texttt{polars}}~\cite{polarsdevelopmentteam_polars_2025}.
We opted to convert the dataset into a nested Python dictionary when computing the component sizes.
The degree distributions were computed using a \texttt{value\_counts} operation in \texttt{polars}, and the component sizes were computed using the \href{https://github.com/bwesterb/py-tarjan}{\texttt{py-tarjan}} Python package. Both of these computations required 256 GB of memory to run successfully. 

Generally speaking, we recommend against using \href{https://networkx.org/}{\texttt{networkx}} to work with the following network; loading the following network with \href{https://igraph.org/}{\texttt{igraph}} or \href{https://graph-tool.skewed.de/}{\texttt{graph-tool}} consumes relatively tractable amounts of memory, but doing the same with \texttt{networkx} used over 500 GB of memory before even loading the full network. We provide example code to load the following network as a \texttt{graph-tool} or \texttt{igraph} object; for those who prefer to use dataframes to work with the data, we also include a copy of the following network in Parquet format. Note that we include functionality for converting follow dates to integer edge attributes (i.e. days since the epoch) as well as code to load the node and starter pack dataset with corresponding conversions of all relevant dates to integers. This minimizes the dataset's memory footprint while still making it possible to work with the temporal aspects of the dataset. 

When working with \texttt{igraph} and \texttt{graph-tool}, we find that loading the edges in large batches provides the best balance of minimizing memory usage and loading the graph in a timely fashion. If the entire graph is loaded as a \texttt{polars} dataframe and then converted into a \texttt{graph-tool} or \texttt{igraph} object, RAM usage also will quickly exceed 500 GB. For this reason, we recommend loading around 5-10\% of the edges at a time, streaming data from the \texttt{csv.gz} file, then inserting them into the graph object. Parsing and inserting 100 million edges consistently took around 6 minutes for the \texttt{graph-tool} loading process. For \texttt{igraph}, adding edges takes longer as the graph grows. Parsing and inserting the first 250 million edges took around 15 minutes, while the last set of edges, numbering less than 250 million, required 35 minutes. Overall, loading the graph as an \texttt{igraph} object used 460 GB of RAM and took 5.5 hours on an HPC instance, while loading the graph into \texttt{graph-tool} consumed 310 GB of RAM and took 2.75 hours.

The starter pack data is a bit more manageable, requiring an order of magnitude less memory.
However, as mentioned in the section on data validation, the line graph computed from the starter pack data is extremely dense; 29\% of all possible pairs of starter packs share at least one member in common.
When working with the edge connection data, one can dramatically decrease the number of shared starter pack members by removing the highest degree accounts.

As noted when discussing the validity of the data, a small percentage of the starter packs and following relationships have incorrect time stamps.
In the case of the starter packs, we recommend either discarding the packs with incorrect dates, setting the times to the creator's account creation date, or sampling an account age at the time of starter pack creation by bootstrapping the distribution of realizable account ages presented in Fig.~\ref{fig:temporal_sp_stats}(b).
In the case of the following relationships, we recommend either discarding the follow events that have impossible dates or setting the follow dates to the maximum of the follower and followed accounts' creation dates. 

\section*{Data Availability}

The dataset is hosted on the Social Media Archive (SOMAR) at the Inter-university Consortium for Political and Social Research (ICPSR) at \href{https://doi.org/10.3886/ICPSR300499}{https://doi.org/10.3886/ICPSR300499}.

\section*{Code Availability}

The code used to analyze the starter pack and following networks is available on GitHub\\
(\href{https://github.com/nwlandry/a-blue-start}{https://github.com/nwlandry/a-blue-start}) and at Ref.~\cite{smith_code_2026}.

\section*{Author Contributions}
A.H.S.: project conception, data acquisition, data processing, data analysis, writing. I.A.: data analysis, writing. S.K.: writing. B.F.W.: project conception, writing. N.W.L.: project conception, data acquisition, data processing, data analysis, writing.

\section*{Acknowledgments}

N.W.L. acknowledges support from the University of Virginia Prominence-to-Preeminence (P2PE) STEM Targeted Initiatives Fund, SIF176A Contagion Science. A.H.S. acknowledges support from the National Science Foundation Graduate Research Fellowship Program under Grant No. 1938052. Any opinions, findings, and conclusions or recommendations expressed in this material are those of the authors and do not necessarily reflect the views of the National Science Foundation. Pacific Northwest National Laboratory is operated by Battelle Memorial Institute under Contract DE-ACO6-76RL01830. PNNL Information Release Number: PNNL-SA-211224.

\section*{Competing Interests}

The authors declare no competing interests.

\bibliographystyle{apsrev4-1}
\setlength{\bibsep}{0pt plus 0.3ex}
{\footnotesize
\bibliography{references}}
\end{document}